\documentclass[secnumarabic, graphics,floatfix, nofootinbib,tightenlines,nobibnotes, aps, prl, 12pt]{revtex4-1}
\usepackage{graphicx}
\usepackage{subfigure}
\usepackage[english]{babel}
\usepackage[utf8]{inputenc}
\usepackage{amsmath,amssymb}
\usepackage{amsfonts}
\usepackage{multirow}
\usepackage{pstricks}
\usepackage{float}
\usepackage{color}
\usepackage{epsfig}

\begin{document}

\title{Centrality and transverse momentum dependence of dihadron correlations in a hydrodynamic model}

\author{Wagner M. Castilho$^{1}$}
\author{Wei-Liang Qian$^{2,1,3}$}

\affiliation{$^1$Faculdade de Engenharia de Guaratinguet\'a, Universidade Estadual Paulista J\'ulio de Mesquita Filho, 12516-410, Guaratinguet\'a, SP, Brazil}
\affiliation{$^2$Escola de Engenharia de Lorena, Universidade de S\~ao Paulo, 12602-810, Lorena, SP, Brazil}
\affiliation{$^3$School of Physical Science and Tecnology, Yangzhou University, 225002, Yangzhou, Jiangsu, P.R. China}

\date{Feb. 16, 2018}

\begin{abstract}
In this work, we study the centrality as well as transverse momentum dependence of the dihadron correlation for Au+Au collisions at 200A GeV. 
The numerical simulations are carried out by using a hydrodynamical code NeXSPheRIO, where the initial conditions are obtained from a Regge-Gribov based microscopic model, NeXuS. 
In our calculations, the centrality windows are evaluated regarding multiplicity.
The final correlations are obtained by the background subtraction via ZYAM methods, where higher harmonics are also considered explicitly. 
The correlations are evaluated for the 0 - 20\%, 20\%-40\% and 60\%-92\% centrality windows. 
Also, the transverse momentum dependence of the dihadron correlations is investigated.
The obtained results are compared with experimental data. 
It is observed that the centrality dependence of the ``ridge" and ``double shoulder" structures is in consistency with the data.
Based on specific set of parameters employed in the present study, it is found that different ZYAM subtraction schemes might lead to different features in the resultant correlations.
\end{abstract}

\maketitle
\section{Introduction}

The strongly interacting matter produced in relativistic heavy-ion collisions results in a new state of the matter, known as Quark-Gluon Plasma (QGP).
Theoretically, the latter is described by Quantum Chromodynamics (QCD) \cite{qcd,qcd1}. 
After the collisions, the system evolves hydrodynamically, cooling and becoming more diluted until it reaches the instant of thermal freeze-out $\tau$, where the system decouples, and the hadronic particles do not interact any more \cite{qgp,jet}. 
The QGP state of matter can be explored by investigating the final hadrons, especially those created in hard processes whose trajectories traverse the entire system \cite{jetEloss}. 

Collisions of Au+Au nuclei have a reaction zone with a transverse diameter of approximately 10 fm.
Subsequently, a hard parton created near the upper edge of the hypersphere and moving inward needs 10 fm/c before emerging from the other side.
This amount of time corresponds to a period sufficient for the hot matter to thermalize, expand, cool and almost attain the decoupling hypersphere. 
Usually, it is assumed that the system reaches the thermal equilibrium at the very beginning of the collision, $t \leq 1$fm/c, where the matter interacts strongly and therefore non-perturbatively.
The pressure against surrounding vacuum leads to a collective expansion that can be treated hydrodynamically. 
The plasma in heavy ion collisions can be explored using various signatures. Among others, two of them are the suppression in the magnitude of high transverse momentum particles in comparison to p+p collisions \cite{corr,cor} and the flow anisotropy of hadrons of small transverse momentum ($p_{T} \leq$ 3-4 GeV/c) \cite{vn,vn2}. 
In fact, the latter is a sensitive signature of the occurrence of plasma.
The flow anisotropy can be quantified in terms of the azimuthal distribution of particles in the momentum space with respect to the reaction plane.
It is attributed to the anisotropy in plasma expansion as a fluid, which is closely related to the transport properties of the QGP \cite{anysotropic2,anysotropic1}.

The particle correlation is also an important observable in the study of QGP.
The measured correlations of high transverse momentum particles of A+A collisions display an interesting structure on the near side of trigger particle, around $\Delta \phi \approx 0$, known as the ``ridge".
The ridge structure receives its nomenclature owing to the observed long extension in $\Delta \eta$ direction \cite{ridge}. 
Such correlation structure has also been observed in pp and p+A collisions at ``Large Hadron Collider (LHC)" \cite{ridge1}. 
Although in general, the production of the ridge varies with the centrality \cite{ridge2,ridge3}, the measurements of p+Pb collisions have shown that the ridge yields corresponding to the jet productions are approximately constant for different centrality windows. 
In opposite side of the trigger particle,  for Au+Au collisions, the correlation structure presents a double peak structure that changes continuously from a double-peak to a single peak when one goes from the central to the peripheral collisions.
The modification in the characteristic of the correlation structure is also observed as a function of the angle of trigger particles $\phi_{s}$, which involves from the double peak in the out-of-plane direction ($\phi_{s} =$  $\pi $/2) to a single peak in the in-plane direction ($\phi_{s} =$ 0) \cite{d-hadr}.
The above phenomenon is not observed in p+p and d+Au collisions \cite{cor}. 

The above-mentioned analysis of dihadron correlation in azimuthal angle is fulfilled by the projection of the measured correlation on the $\Delta \phi$ axis.
It consists of two contributions: that comes from hard process $C(\Delta \phi)$ which is associated with the hard trigger particles and, that relates to the azimuthal anisotropic distribution of soft hadrons which is attributed to the flow anisotropy owing to soft scatterings. 
It is well known that the event by event fluctuating initial conditions (IC) play an essential role in the observed correlation structure in A+A collisions.
Using NeXuS event generator, tubular structure arises mainly from hard parton scattering. 
It is the hydrodynamical evolution that transforms these structures into the ridge.
However, it is noted that the ridge and double peak were not originally understood to be a collective nature of the system, and they were attributed to jet quenching either by Cherenkov effects \cite{chkov,chkov1} and/or by Mach cone \cite{mach,mach1}.
In our previous studies, the origin of the double peak in the away-side \cite{tube} is attributed to the local deflection of the fluid by the hot-spots, which effectively gives rise to the third harmonic coefficient $v_{3}$.
Hydrodynamical studied using IC with tubes positioned in the peripheral region of the system showed that they are mainly responsible for the observed features in dihadron correlations in A+A collisions \cite{hotspt,hotspt1}.

The paper is organized as follows.
In the next section, we discuss the NeXSPheRIO model and the calculation setup, in particular, the classification of centrality windows adopted in this work.
Additionally, the properties of the event by event fluctuations are analyzed.
The results on dihadron correlations and discussions are given in Section III.
The correlations are calculated for three centrality windows, namely, 0 - 20\%, 20\%-40\%, and 60\%-92\%. 
Also, the transverse momentum dependence of the dihadron correlations is studied.
The obtained results are compared with experimental data. 
Concluding remarks are given in the last section.

\section{The NeXSPheRIO model and calculation setup}

\subsection{Fluctuating initial conditions}

In our study, the IC for the hydrodynamic evolution are provided by an event generator, which describes the complex microscopic process of interactions between partons during the initial stage of the collisions.
We make use of NeXuS \cite{Nexus}, which is based on Regge-Gribov field theory \cite{R-Grib}. 
The IC are expressed in terms of the energy-momentum tensor, which in turn is used to construct the four-velocity of the fluid $u^{\mu}$ and the conserved currents $n^{i}$ in the initial proper time $\tau \approx$ 1fm \cite{bjor}.
Therefore, although NeXuS produces jets, they ``melts" and merge into the thermalized media before fed to the hydrodynamic model since the latter assumes that the system is locally thermalized and the energy-momentum tensor is diagonalized.
Subsequently, the system undergoes a collective evolution described by SPheRIO \cite{ebye2,sph5} which is a three-dimensional (3D) ideal hydrodynamic code. 
This code that unites the two models is called NeXSPheRIO.

\begin{figure}[h!]
\begin{center}
\includegraphics[width=5cm, height=5cm]{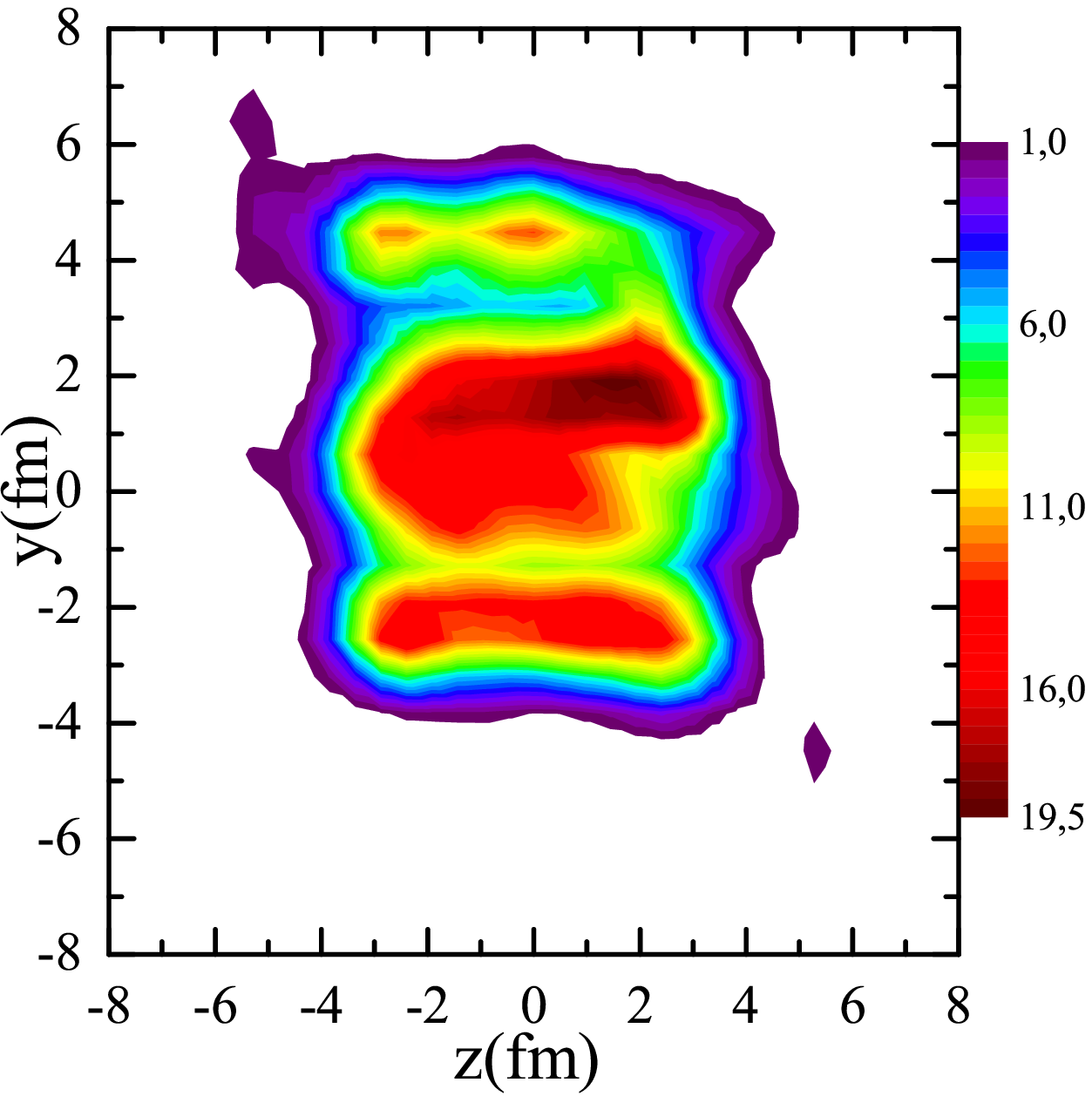} %[$\eta=0$]
\ \ \ \ \
\includegraphics[width=5cm, height=5cm]{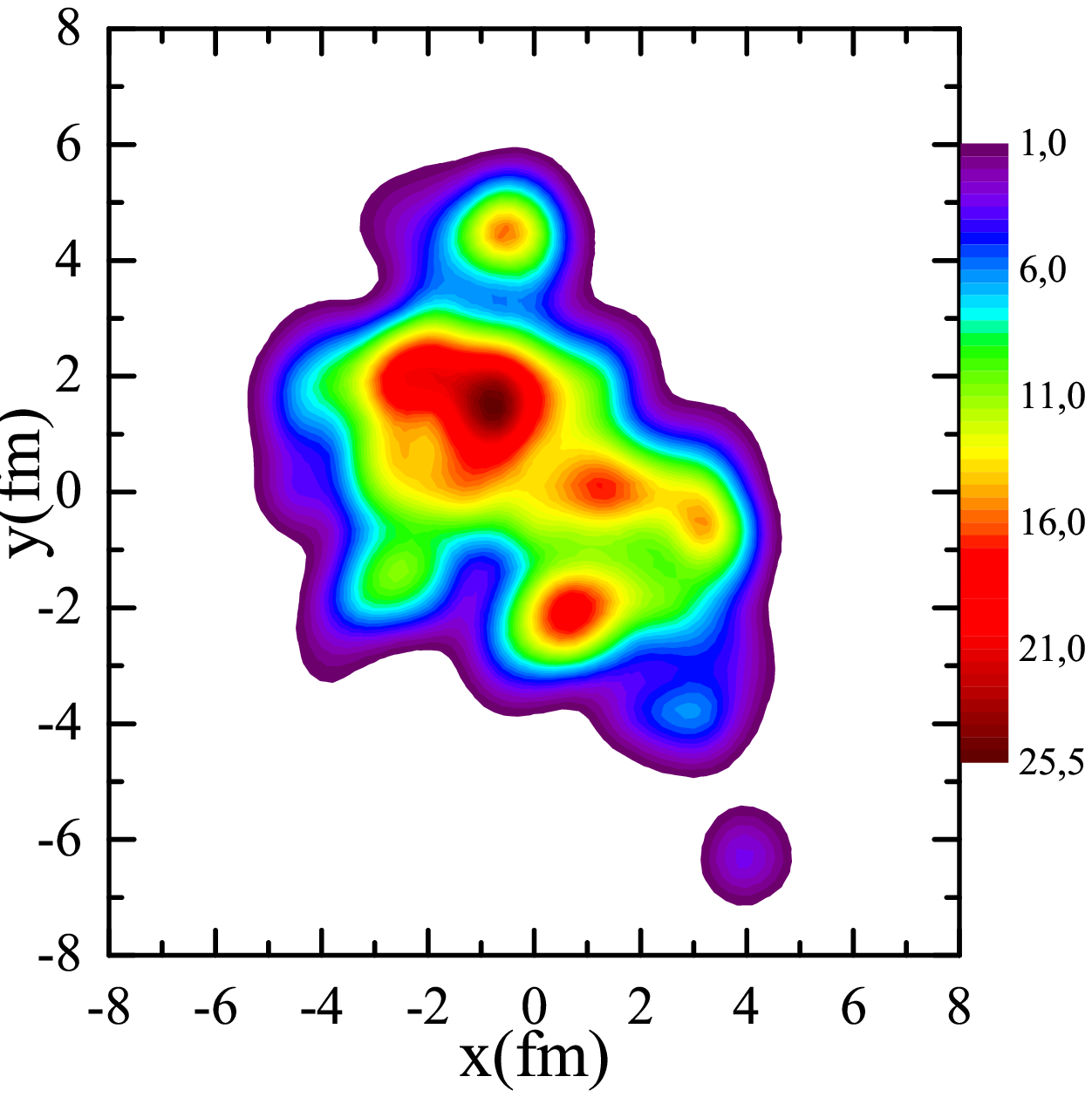} %[$x=0$]
\end{center}
\caption{(Color online) Energy density generated by NeXuS in the centrality window of 0-10\% for 200A GeV Au+Au collisions.}
\label{ic}
\end{figure}

The hadron distributions, as well as dihadron correlations, are evaluated in terms of the hadrons after they freely reached the detectors in the laboratory after a sufficiently long time $\ge 10 $fm/c.
An essential physical content which sensitively affects the observed dihadron correlations is the fluctuating IC \cite{ebye2,ebye1}. 
In Fig.\ref{ic}, we show the initial energy density in the centrality window of 0-10\% generated by NeXuS for 200 AGeV Au+Au collisions.
The right panel displays a cross-sectional view of the pseudorapidity $\eta$=0 and the same IC is presented for $x$ = 0 in the left panel. 
It is noted that longitudinal tubes are observed in the IC.

These high-energy tubular structures, related to the soft partons or strings \cite{Nexus}, can not be treated by QCD perturbatively. 
The properties of the tubular structure, such as energy density, diameter, and position, vary randomly from event by event.
As shown in Table \ref{ci}, all these quantities decrease as one goes from central to peripheral collisions. 
The calculations were carried out by using 1000 fluctuating events randomly generated by NeXuS for each of the eight centrality windows, starting from the most central 0 - 10 \% to the most peripheral 80 - 92 \% for Au+Au collisions. 
The average volume of the tubes was estimated by the fitting the boundary of the cross-section area of the tube at $\eta=0$ to an ellipse and calculating the average height of the tube from individual events as shown in Fig.\ref{ic}.
As event by event IC play an essential part in the study of dihadron correlations, they will be employed in the present study.

\begin{table}[h!]
\centering
\caption{Analysis of fluctuating IC in terms of tubes. 
The average values are obtained by using 1000 events generated by NeXuS for 200A GeV Au+Au collision in different centrality windows.}
\vspace{0.5cm}
\begin{tabular}{|c|c|c|c|c|}
\hline
Collision & Variation of  & Average number & Average volume & Average energy density \\ 
centrality (\%)& impact parameter $b$ (fm) & of tubes & of tubes & of tubes \\ \hline
 00 - 10 & 4.783 & 3.37 & 53.62 & 31.02 \\
 10 - 20 & 1.982 & 2.91 & 53.03 & 30.61 \\
 20 - 30 & 1.521 & 2.62 & 49.84 & 29.23 \\ 
 30 - 40 & 1.282 & 2.33 & 49.62 & 27.43 \\
 40 - 50 & 1.129 & 2.11 & 49.70 & 26.17 \\
 50 - 60 & 1.021 & 1.87 & 47.01 & 22.97 \\
 60 - 80 & 1.813 & 1.55 & 35.69 & 15.43 \\
 80 - 92 & 0.979 & 1.20 & 33.38 & 10.85 \\ \hline
\end{tabular}
\label{ci}
\end{table}

\section{Collision centralities}

According to the Glauber model, the centrality is a geometric parameter defining the region of the interpenetration of the incident nuclei. 
The physical measurements in the laboratories, such as the total number of particles, particle spectrum \cite{pt}, two-particle correlation \cite{cor}, among others, are closely related to this parameter.
In practice, the impact parameter, $b$, is often associated with other quantities such as the number of participant nucleons, number of binary collisions, number of charged particles produced in an event. 

\begin{figure}[h]
\begin{center}
\subfigure[]{
\includegraphics[height=5cm]{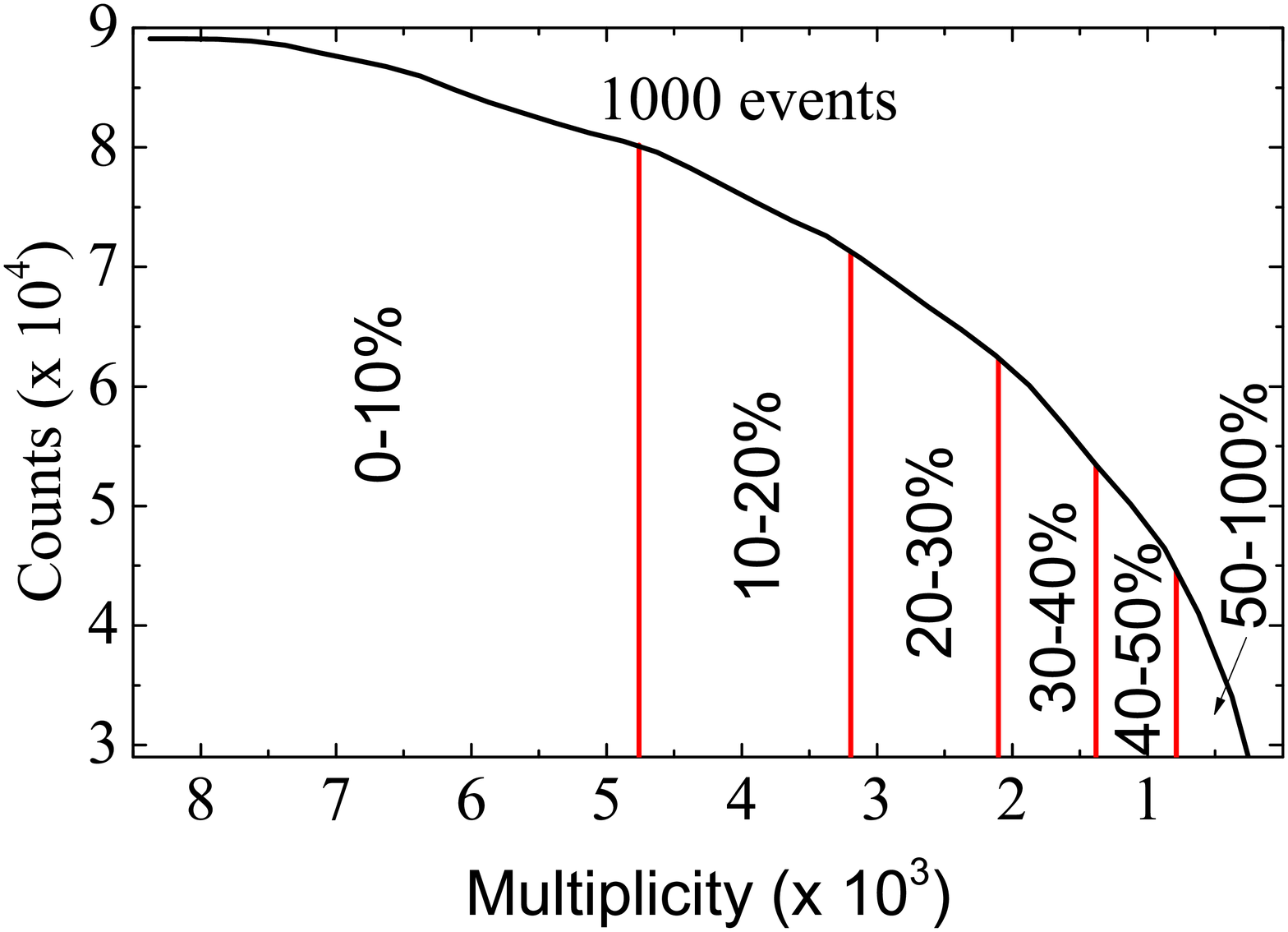} %, height=7cm para altura
}
\subfigure[]{
\includegraphics[height=5cm]{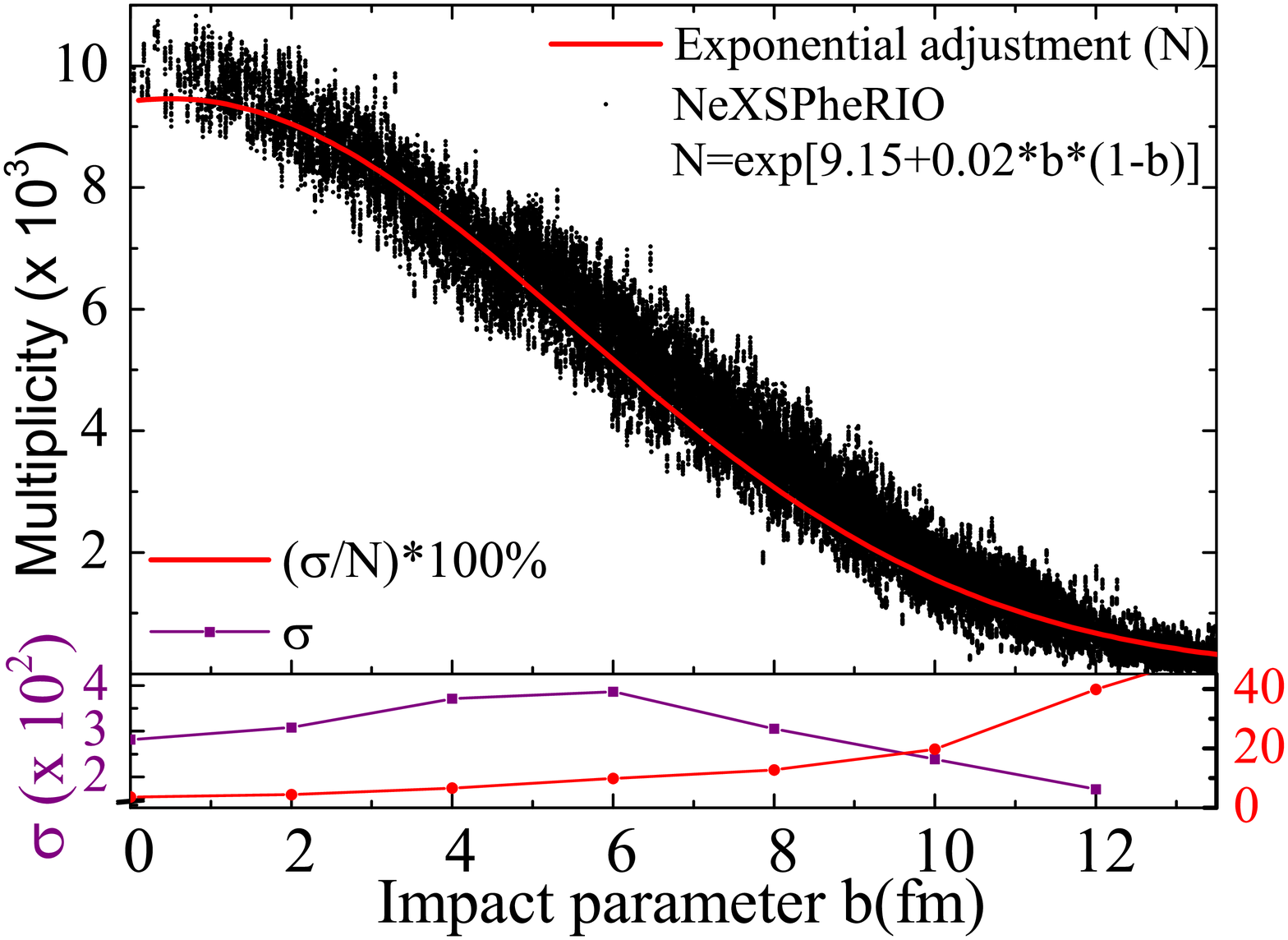}
}
\end{center}
\caption{(Color online) (a) Centrality calculations of collisions using NeXSPheRIO, as defined by the experimentalists (b) The relationship between multiplicity and the impact parameter $b$.}
\label{sj}
\end{figure}

A standard method to define a centrality window by the experimentalists is to evaluate the total number of particles produced in an event, known as multiplicity.
On average, it is proportional to the number of participating nucleons, which in turn is inversely proportional to the impact parameter \cite{b}. 
In this representation we have: 10\% of collisions with the largest multiplicities correspond to the 0 - 10\% centrality window, while the 20\% collisions of the largest multiplicities with the previous ones excluded correspond to the centrality window of 10\% - 20\%, and so forth the other centrality windows are defined. 
The centrality of collision as a function of multiplicity is shown in Fig.\ref{sj}a.
The relationship between multiplicity and impact parameter $b$ is shown in Fig.\ref{sj}b. 
The calculations are performed by employing NeXSPheRIO, where 1000 events are generated for 200A GeV Au + Au collisions. 
The black diamonds shown in Fig.\ref{sj}b were obtained by NeXSPheRIO and the solid red curve represents the least-squares fitting to an exponential function. 
We note that for a given impact parameter, $b$ = 8 fm for example, the multiplicity has a deviation of 12.37\% ($\langle N_{p} \rangle \pm$ $\sigma$ = 121 $\pm$ 14.97), which is due to event by event fluctuations. 
The standard deviation of the multiplicity ($\sigma$) varies with the impact parameter, it reaches the maximum in the mid-central region ($b = 5-8$ fm), as observed by the filled black squares in the lower part of Fig.\ref{sj}b. However, the relative variation ($\sigma/\langle N_{p}\rangle \times$ 100\%) increases monotonically as one considers more peripheral windows, as shown by the filled red circles.

\begin{figure}[h]
\begin{center}
\includegraphics[width=8cm]{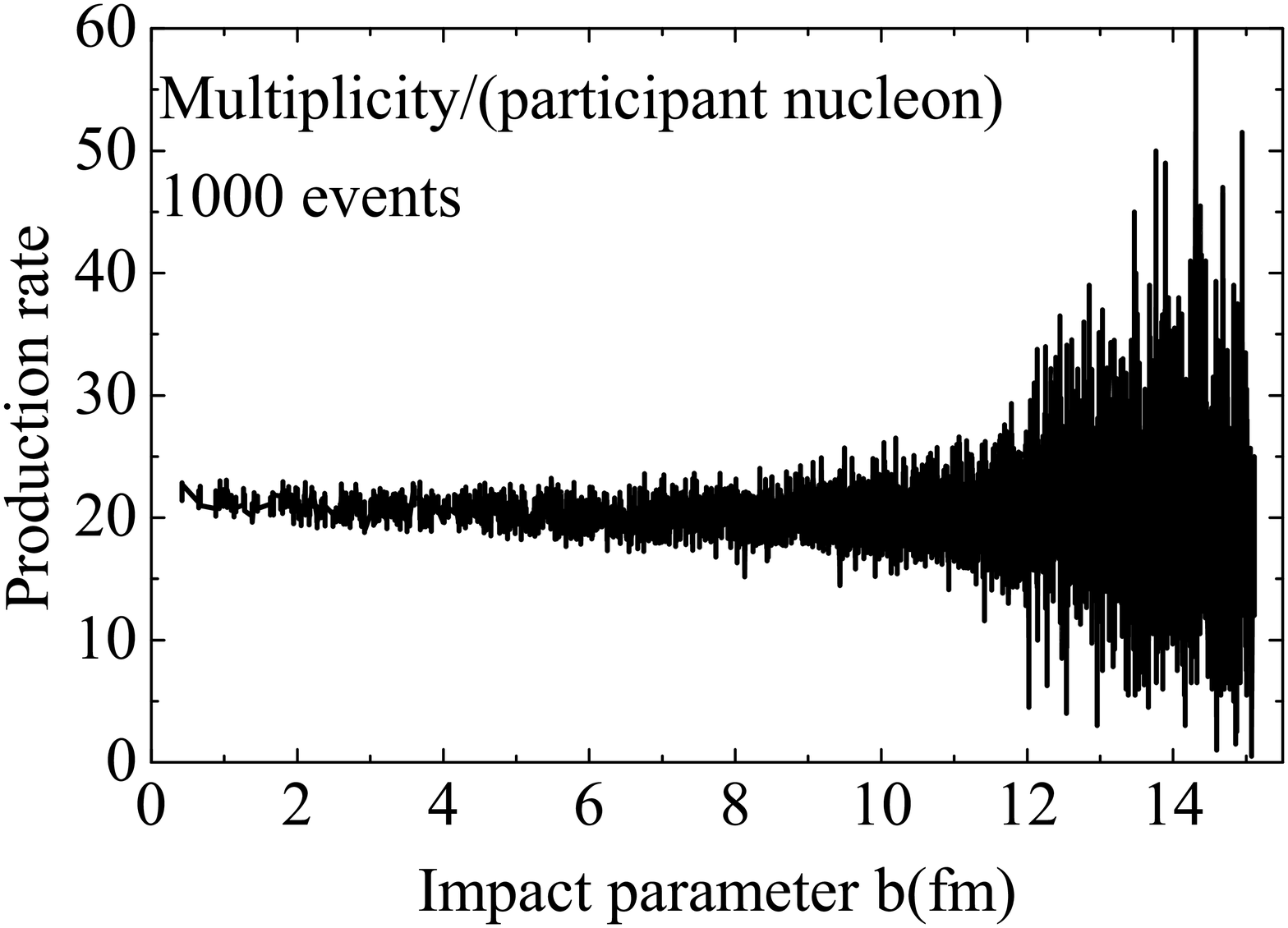} %, height=7cm para altura
\includegraphics[width=8cm]{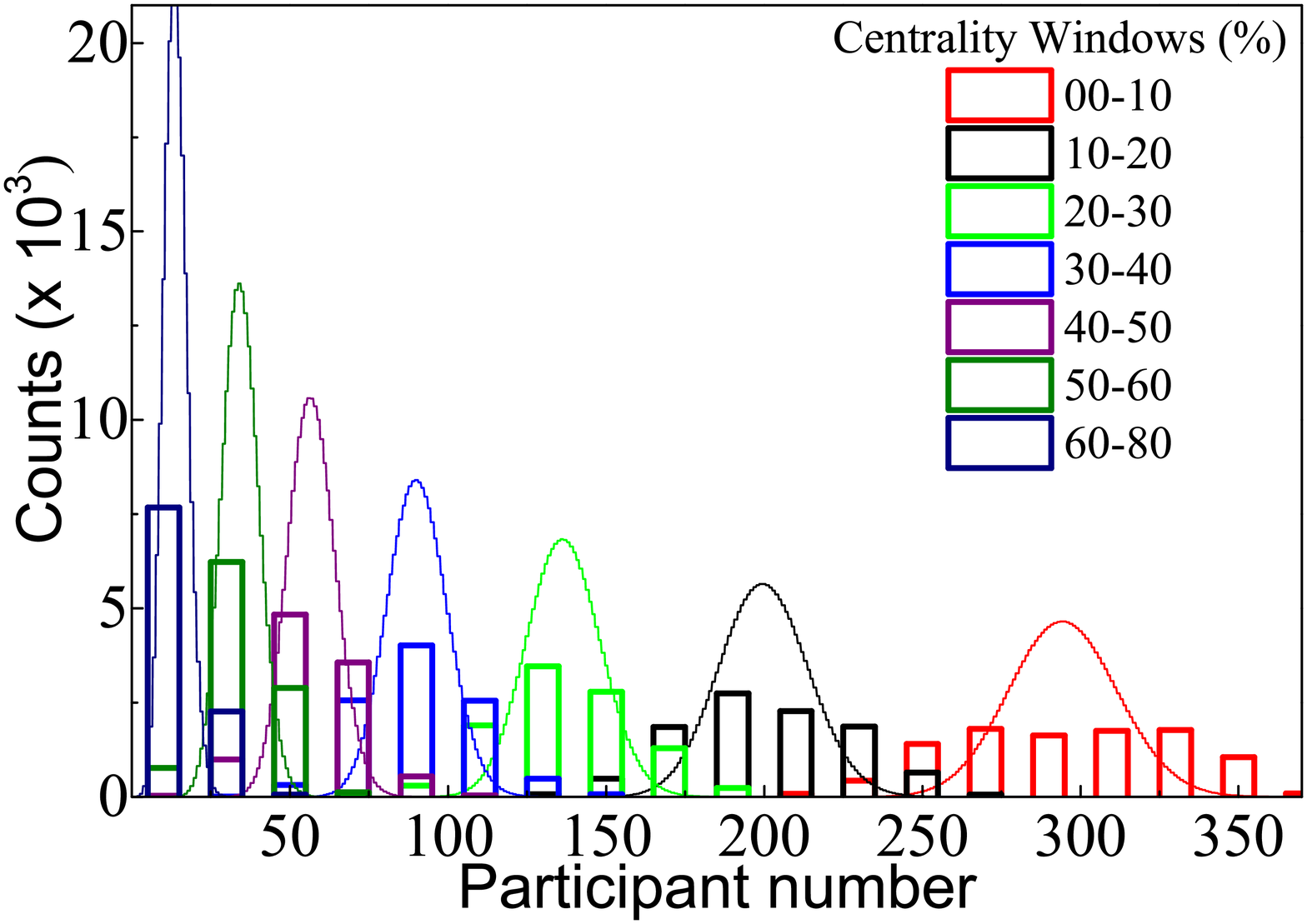} %, height=7cm para altura
\end{center}
\caption{(Color online) Particles produced for different centralities. Right: the ratio between the number of produced hadrons to the number of participants as a function of impact parameter. 
Left: histogram of the number of participants for different centrality windows.}
\label{npxmult}
\end{figure}

In Fig.\ref{npxmult}a, we show the ratio between the produced hadrons to the number of participants as a function of impact parameter. 
The distribution of the number of participants is presented in Fig.\ref{npxmult}b in terms of the histogram for different centrality windows, together with the fittings to Poisson distributions.
The average number of hadrons produced by each participant nucleon is around 20, and the value seems to be independent of the impact parameter. 
However, the uncertainty of the ratio increases for more peripheral windows Fig.\ref{sj}b. 
On the other hand, the number of participants increases as one considers more peripheral collisions, as expected.
However, the standard deviation of the number of participants also increases for more peripheral windows, which is probably related to the observed increase of standard deviation in Fig.\ref{npxmult}a.

In the present study, we adopt the more realistic classification for the centrality windows, where the centrality are defined in terms of multiplicities.

\section{Dihadron azimuthal correlations}

\subsection{Centrality dependence of dihadron correlations}

In this section, the azimuthal angle correlations between hadron pairs (dN/d$\Delta \phi$) are analyzed.
The range of pseudorapidity $ |\eta| < 1$ is considered, in accordance to the experimental setup.
In a previous study, the calculations of dihadron correlations as a function of the azimuth angle of the trigger particles relative to the plane of event $\phi_{s} = | \phi_{trig} - \psi_{r}|$ was carried out by the code NeXSPheRIO \cite{hotspt}. 
The results were compared with the experimental data obtained by the PHENIX collaboration \cite{cor}. 
By using the same set of parameters, the centrality dependence of the correlations is carried out for central windows 0 - 20\%, 20\% - 40\%, and 60\% - 92\%. 
To reproduce the experimental procedure and obtain a good statistic, the centrality windows were equally divided into two smaller classes, for central (0 - 10\% and 10\% -20\%), mid-central (20\% - 30\% and 30\% - 40\%) and peripheral (60\% - 80\% and 80\% - 92\%) windows respectively.
A total of 3,500 events are generated for the central, 2,000 for the mid-central and 5,000 for the periphery windows. 
For each initial condition, the Monte Carlo hadron generator was invoked for 200 times. 
The analyzed trigger particles were taken from transverse momentum range $2 < p_{T}^{trig}< 3 $GeV/c and the associated particles from $0.4 < p_{T}^{ass} < 1 $ GeV/c. 
The calculated correlations between trigger and associated particles $C_{\text{prop}}(\Delta \phi, \Delta \eta)$ are then projected onto the axis of the azimuthal angle difference $\Delta \phi$ to obtain $C_{\text{prop}}(\Delta \phi)$.
It is understood that the one-particle distribution of the associated particles is mostly determined by the anisotropic flow, which is predominantly generated by the elliptic flow $v_{2}$.
Such contribution therefore can be subtracted by the ZYAM method $C_{\text{ZYAM}}(\Delta \phi)$, according to the procedure adopted by the PHENIX Collaboration  
\begin{equation}
C_{\text{ZYAM}}(\Delta \phi) = B \left(1+2v_{2}^{ass}v_{2}^{trig}\cos2\Delta\phi \right)    \label{zya1}
\end{equation}
where $B$ is background normalization, $v_{2}^{ass}$ ($v_{2}^{trig}$) is the second harmonic coefficient of the associated (trigger) particles with respect to the event plane. 
The elliptical flow coefficients $v_{2}$ are calculated by the event plan method, and their respective values are shown in Table \ref{vnt}. 

\begin{table}[h]
\centering
\caption{The elliptic flow $v_{2}$, triangular flow $v_{3}$ and quartic flow $v_{4}$ obtained by the event plan method for different centrality windows.}
\vspace{0.5cm}
\begin{tabular}{|c|ccc|ccc|ccc|}
\hline
range & \multicolumn{3}{|c|}{central window} & \multicolumn{3}{|c|}{mid-central window} & \multicolumn{3}{|c|}{peripheral window} \\ 
 $p_{T}$ (GeV)& $v_{2}$ & $v_{3}$ & $v_{4}$ & $v_{2}$ & $v_{3}$ & $v_{4}$ & $v_{2}$ & $v_{3}$ & $v_{4}$ \\ \hline
 0.4 - 1.0 & 0.0399 & 0.0151 & 0.0068 & 0.0729 & 0.0212 & 0.0097 & 0.0739 & 0.0210 & 0.0123 \\
 1.0 - 2.0 & 0.0939 & 0.0457 & 0.0271 & 0.1543 & 0.0600 & 0.0373 & 0.1460 & 0.0528 & 0.0447 \\
 2.0 - 3.0 & 0.1387 & 0.0938 & 0.0675 & 0.2466 & 0.1156 & 0.0878 & 0.2369 & 0.1029 & 0.1153 \\ \hline
\end{tabular}
\label{vnt}
\end{table}

\begin{figure}[h!]
\begin{center}
\includegraphics[width=15cm]{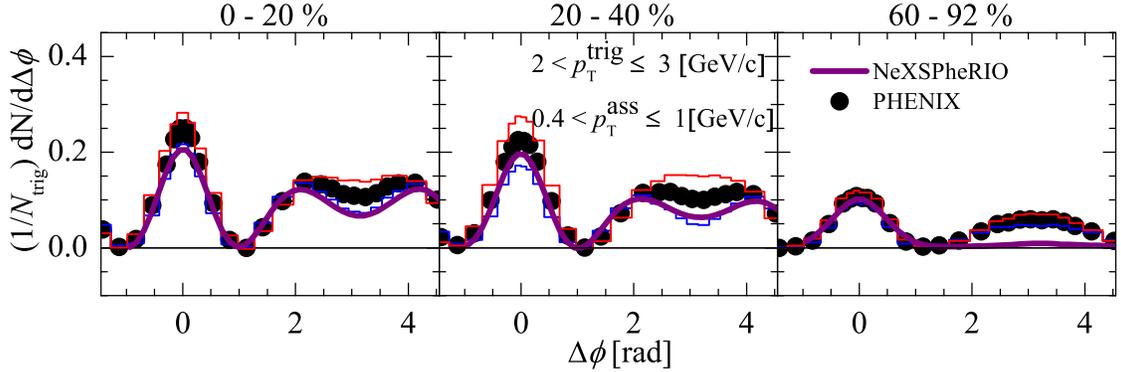}
\end{center}
\caption{(Color online) Calculated dihadron correlations by using the NeXSPheRIO. 
The resultant correlations are corrected by the ZYAM method where the subtraction of the elliptical flow is considered.
The NeXSPheRIO results are represented by the solid purple curves. 
The experimental data \cite{cor} are shown by black filled the circles, and the red/blue solid curves above and below the data represent its uncertainty. 
The transverse momentum range $|\eta|$ $<$ 1, $2 \le p_{T}^{trig} \le 3$, $0.4 \le p_{T}^{ass} \le 1$ are used in accordance with the experimental data.}
\label{corjan}
\end{figure}

The results of correlations are given in Fig.\ref{corjan}.
It shows one peak in the away-side for peripheral collisions 60\% - 92\%, as well as a broader peak on the near-side. 
The structure continuously changes to a double peak structure for mid-central (20\% - 40\%) and central (0\% - 20\%) collisions. 
The correlation decreases considerably for peripheral collisions, as shown in Fig.\ref{sj}, together with the overall multiplicity. 
These resulting correlations are in consistency with our previous findings \cite{hotspt}.
Therefore, we conclude that the modified classification of centrality windows does not affect the dihadron correlations qualitatively.

\subsubsection{Transverse momentum dependence of dihadron correlations}

\begin{figure}[h!]
\begin{center}
\subfigure[][]
{
\includegraphics[width=15cm]{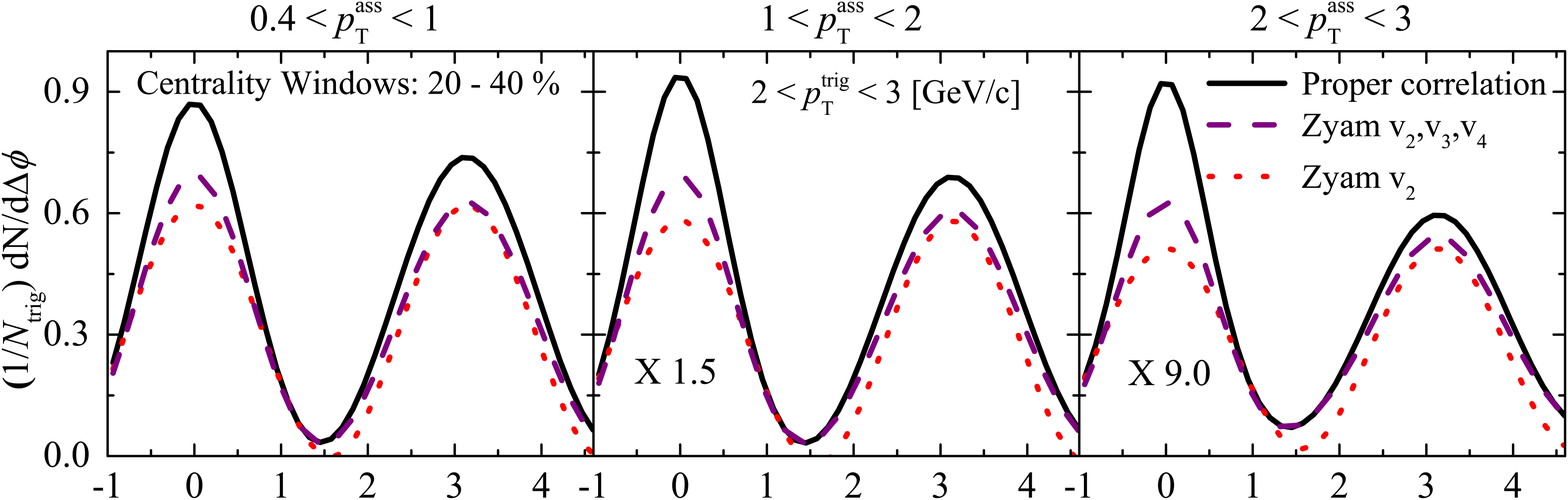}
}
\par
\subfigure[][]
{
\includegraphics[width=15cm]{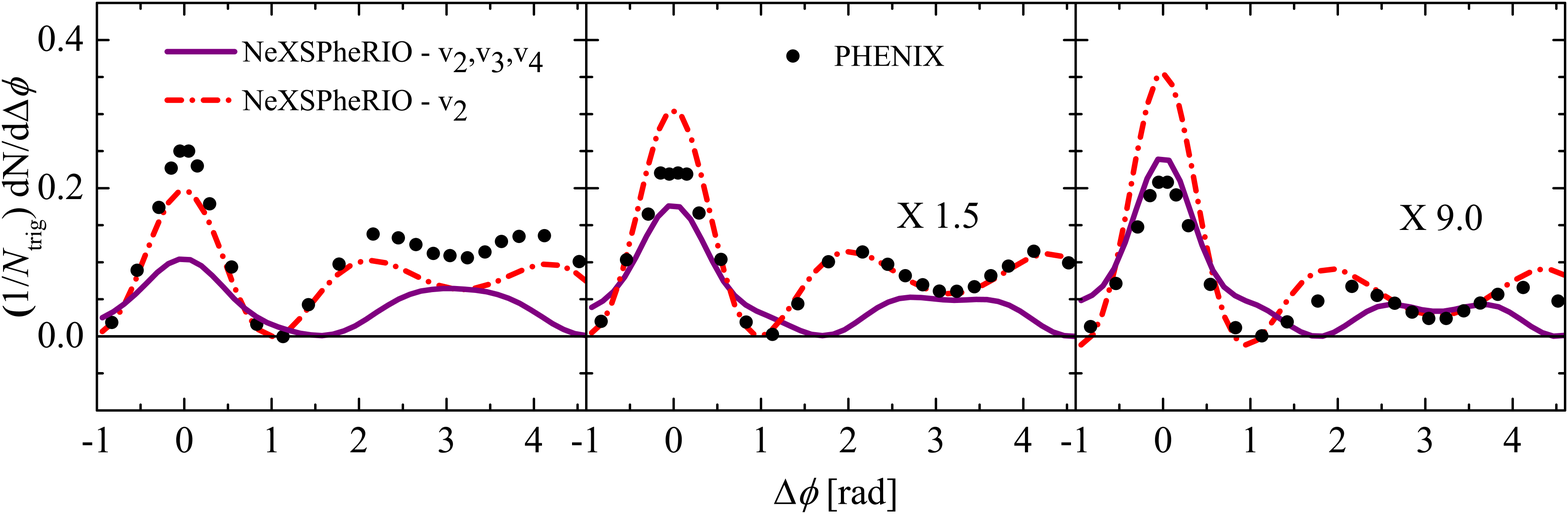}
}
\par
\end{center}
\caption{(Color online) Calculated correlations by using two different ZYAM methods. 
The results obtained by subtraction of harmonics $v_{2}$, $v_{3}$ and $v_{4}$ are shown in solid purple curves.
Those by subtraction of $v_2$ are shown in red dash-dot curves and compared with the experimental data by PHENIX \cite{cor} in filled black circles. 
(a) The proper correlation as well as the background by using two different ZYAM methods. (b) The resultant correlations.}
\label{zyamv2v3v4}
\end{figure}

In this section, the $p_{T}$ dependence in azimuthal correlation were analyzed for the mid-central windows 20\% - 40\%.
Again, the particles are only taken from the pseudorapidity range $| \eta |< 1 $.
To compare with the experimental data obtained by the PHENIX collaboration \cite{cor}, the trigger particles are from the interval $2 < p_{T}^{trig} < 3$ GeV/c and the associate particles are from three different intervals $0.4 < p_{T}^{ass} < 1$, $1 < p_{T}^{ass} < 2$ and $2 < p_{T}^{ass} < 3$ GeV/c. 
It is found that the calculated correlations decrease considerably as the transverse momenta of the associated particle increase.
The correlations for $2 < p_{T}^{ass} < 3$ GeV/c is approximately 9 $\times$ smaller in comparison to the correlation for $0.4 < p_{T}^{ass} < 1$, as shown in Fig.\ref{zyamv2v3v4}. 
The calculated results, with the subtraction of elliptical flow only, slightly overestimate the data as the transverse momenta of the associated particles increase. 

We may also take into account additional contributions from higher order harmonic coefficients, namely, triangular flow $v_{3}$ and quartic flow $v_{4}$.
In this case, the contribution from the background becomes \cite{d-hadr}
\begin{equation}
C_{\text{ZYAM}}(\Delta \phi) = B \left(1+2v_{2}^{ass}v_{2}^{trig}\cos2\Delta\phi+2v_{3}^{ass}v_{3}^{trig}\cos3\Delta\phi+2v_{4}^{ass} v_{4}^{trig} \cos4\Delta\phi \right)    \label{zya1}
\end{equation}
where $B$ is the background normalization. $v_{2}^{ass}$, $v_{3}^{ass}$ and $v_{4}^{ass}$ ($v_{2}^{trig}$, $v_{3}^{trig}$ and $v_{4}^{trig}$) are the second, third and quartic harmonic coefficients of the associated (trigger)particles. 
The harmonic coefficients are calculated by event plan method and shown in Table \ref{vnt}.
It is noted that the coefficients of the anisotropic flow increase as one considers more peripheral collisions or larger transverse momentum.
They reach the maxima in the mid-central window.
The magnitude of flow coefficient decreases with the increase of harmonic order. 

Besides the contributions from the elliptic flow, the $v_{3}$ term creates a structure of three peaks located at $\Delta \phi = 0$, which contributes to the ridge in the ``near-side'' and, others two peaks in $\Delta \phi$ = 2$\pi$/3 and 4$\pi$/3. 
The latter strengthen the double peaks on the ``away-side'' opposite to the direction of the trigger particle.
Although the major contribution to the background in the ZYAM method still comes from the elliptic flow, we note that for this specific setup, the contributions from higher order harmonics significantly modify the resulting correlation yields.
Therefore, in order to properly subtract the flow background, it seems to be meaningful to subtract the contributions of all three flow harmonics.
This is shown in Fig.\ref{zyamv2v3v4}.
In Fig.\ref{zyamv2v3v4}a, we present the contributions of the background $C_{\text{ZYAM}}(\Delta \phi)$ as well as the proper correlations $C_{\text{prop}}(\Delta \phi)$ when considering three harmonic coefficients. 
The results of the subtraction of $v_{2}$, $v_{3}$ and $v_{4}$ are shown in solid purple curves. 
Those with only the subtraction of the elliptic flow $v_{2}$ are represented by red dash-dot lines together with the published PHENIX data \cite{cor} in filled black circles.  
In Fig.\ref{zyamv2v3v4}a, one observes that the background contribution is mostly dominated by the contribution from $v_2$ and therefore the difference in the ZYAM background between the two methods of subtraction does not seem to be too significant.
However, the background with only the subtraction of $v_{2}$ has a slightly broader peak on the away-side.
As a result, the subsequent subtraction leads to resultant double peaks on the away-side of the correlation.
As shown in Fig.\ref{zyamv2v3v4}b, the hydrodynamical results agree reasonably well with the data.
On the other hand, it is found that the two methods show a noticeable difference in the resultant correlation.
In particular, in the case where $0.4 < p_{T}^{ass} < 1$, double peaks are observed by using only the subtraction of $v_2$ while a single peak is formed when $v_3$ and $v_4$ are also subtracted.
It is also observed that the away-side structure has its maxima at $\Delta \phi \approx 2\pi/3$ and $4\pi/3$, which implies that the double peak structure is mostly generated by the third Fourier component for the momentum intervals studied in the present work. 
Though the higher order harmonic subtraction was carried out in a recent study concerning the event plane dependence of dihadron correlations \cite{d-hadr}, it is noted there is no data for the particular momentum cut employed in this study. 
In this context, we note that such noticable difference owing to different subtraction methods might be interesting and utilized to descriminate different data-analysis methods.

\section{Concluding remarks}

According to the peripheral tube model \cite{sph-corr-2}, the contribution of the event by event fluctuating IC is manifested by the randomized location of the tubes, but only those tubes in the peripheral regions significantly contribute to the correlations. 
The properties of the correlation structure and its centrality, as well as event plane dependence, are attributed to the collectivity of the system.
In addition to the previous findings \cite{hotspt,hotspt1,tube,sph-vn-4}, the present study provides a reasonable description of the data.
We understand the existing data on dihadron correlations can be reasonably explained in a hydrodynamic framework without explicitly considering the interaction between the jet with the medium. 

In this work, studies of the dihadron correlations of 200A GeV Au+Au collisions were carried out by using a hydrodynamic approach.
The qualitative results of the observed centrality dependence of dihadron correlation do not change by using a different classification of centrality windows.
To investigate the transverse momentum dependence of correlations, we employed two different ZYAM subtractions for the background contribution. 
The analysis carried out by the STAR collaboration \cite{star-plane1,star-plane2,d-hadr} show that the event plane dependence of dihadron correlation is essentially not affected by the subtractions of higher order harmonics, which is subsequently reproduced by a hydrodynamic approach \cite{Castilho:2017ygw}.
There, it was found that the contributions from higher order harmonics do not qualitatively alter the features presented in the resultant correlation yield.
In the present setup, however, it is found that different subtraction schemes may indeed lead to notacible difference, especially for the away-side correlations.
Insterestingly, a Reaction Plane Fit method was proposed~\cite{zyam-rpf-1} recently and employed to estimate the correlation functions in the background dominated region on the near-side~\cite{zyam-rpf-2}.
The resulting correlation does not show the double peak on the away side, neither any dramatic shape modification as a function of centrality.
Subsequently, the authors conclude that the Mach cone is an artifact of the background subtraction and the jets do not fully equilibrate with the medium.
These results further indicate that the effect of the jet in the di-hadron correlation is indeed a subtle subject.
It is therefore understood that caution must be taken in the analysis of di-hadron correlations, especially with the contributions of higher order harmonics in the ZYAM approach. 
Further comparison between different subtraction scheme may provide a better understanding of the underlying physics of observed correlation yields as well as serve to discriminate between different theoretical interpretations.

\section*{Acknowledgments}
We are thankful for valuable discussions with Yogiro Hama and Takeshi Kodama.
We gratefully acknowledge the financial support from
Funda\c{c}\~ao de Amparo \`a Pesquisa do Estado de S\~ao Paulo (FAPESP),
Funda\c{c}\~ao de Amparo \`a Pesquisa do Estado do Rio de Janeiro (FAPERJ),
Conselho Nacional de Desenvolvimento Cient\'{\i}fico e Tecnol\'ogico (CNPq),
and Coordena\c{c}\~ao de Aperfei\c{c}oamento de Pessoal de N\'ivel Superior (CAPES).
This research is also supported by the Center for Scientific Computing (NCC/GridUNESP) of the S\~ao Paulo State University (UNESP).

\bibliographystyle{h-physrev}
\bibliography{ref_castilho}{}

\end{document}